# Feasibility Study of Aerocapture at Mars with an Innovative Deployable Heat Shield

G. Isoletta,[*] and M. Grassi[†]
*University of Naples "Federico II", Naples, 80125, Italy*

E. Fantino[‡]
*Khalifa University of Science and Technology, Abu Dhabi, P.O. Box 127788, United Arab Emirates*

D. de la Torre Sangrà[§]
*Polytechnic University of Catalonia (UPC), Terrassa, 08222, Spain*

and
J. Peláez Álvarez[**]
*Technical University of Madrid (UPM), Madrid, 28040, Spain*

**Performing orbital insertion around Mars using aerocapture instead of a propulsive orbit insertion manoeuvre allows to save resources and/or increase the payload mass fraction. Aerocapture has never been employed to date because of the high uncertainties in the parameters from which it depends, mainly related to atmospheric density modeling and navigation errors. The purpose of this work is to investigate the feasibility of aerocapture at Mars with an innovative deployable drag device, whose aperture can be modulated in flight, and assess the effects of the main uncertainties on the success of the manoeuvre. This paper starts with the presentation of a parametric bi-dimensional analysis of the effectiveness of aerocapture, for which a wide range of uncertainty levels in the atmospheric density and the ballistic coefficient are considered. Then, an application to a real mission scenario is carried out including the error of the targeting manoeuvre performed at the limit of the sphere of influence of the planet. The analyses show the strong influence of the uncertainties in the atmospheric density and the ballistic coefficient, which significantly narrow the solution space**

---

[*] PhD student, Dept. of Industrial Engineering, P.le V. Tecchio, 80; giorgio.isoletta@unina.it.
[†] Full Professor, Dept. of Industrial Engineering, P.le V. Tecchio, 80; michele.grassi@unina.it.
[‡] Assistant Professor, Dept. of Aerospace Engineering; elena.fantino@ku.ac.ae.
[§] Research Assistant, Physics Department, ESEIAAT, C/ Colom 11; david.de.la.torre.sangra@upc.edu.
[**] Full Professor, ETSIAE, Space Dynamics Group, Pl. Cardenal Cisneros, 3; j.pelaez@upm.es.

**and limit its continuity. However, viable solutions for aerocapture can still be identified even in the worst conditions.**

## Nomenclature

| | | |
|---|---|---|
| $\mu$ | = | standard gravitational parameter, m³/s² |
| $r$ | = | magnitude of position vector, m |
| $a_D$ | = | atmospheric drag acceleration, m/s² |
| $\rho$ | = | atmospheric density, kg/m³ |
| $\beta$ | = | ballistic coefficient, kg/m² |
| $v$ | = | magnitude of relative velocity, m/s |
| $m$ | = | mass, kg |
| $C_D$ | = | drag coefficient |
| $A$ | = | cross section area, m² |
| $\theta$ | = | angle of misalignment, ° |
| $J_2$ | = | second-degree term of the mass distribution of Mars |
| $a_{J2}$ | = | acceleration due to J2, m/s² |
| $R$ | = | Mars' radius, m |
| $\Phi$ | = | geocentric latitude of the spacecraft with respect to Mars, ° |
| $a_{Sun}$ | = | acceleration due to the third-body perturbation of the Sun, m/s² |
| $\mu_{Sun}$ | = | standard gravitational parameter of the Sun, m³/s² |
| $R_{Sun}$ | = | position vector of the Sun with respect to Mars' center, m |
| $v_\infty$ | = | hyperbolic excess velocity, km/s |
| $h_\pi$ | = | periapsis altitude, km |
| $\Delta V$ | = | magnitude of the manoeuvre, m/s |
| $\lambda$ | = | longitude angle of the direction of the manoeuvre, ° |
| $\delta$ | = | latitude angle of the direction of the manoeuvre, ° |
| $h_\alpha$ | = | apocenter altitude, km |
| $e$ | = | eccentricity |

# I. Introduction

The lasting scientific interest in Mars exploration and the recently renewed plans for large-scale unmanned missions to the red planet have led to several studies on the feasibility of aerocapture for orbit insertion with aerodynamic decelerators, such as deployable or inflatable drag devices [1–4]. Aerocapture is an aero-assisted orbital maneuver which exploits the drag generated during a single atmospheric passage to decelerate a probe and transfer it from a hyperbolic trajectory to an elliptical orbit. For missions to planets with an appreciable atmosphere, the benefit of this technique lies in its intrinsic time and propellant savings [5]. Since the thermal shielding weighs less than the propellant needed to perform an orbit insertion manoeuvre, opting for aerocapture leads to a larger payload mass and, therefore, a higher scientific return [6].

The traditional Mars Orbit Insertion (MOI) propulsive maneuver inserts the spacecraft into a highly elliptical orbit around Mars. The target orbit is then reached by means of additional burns or with aerobraking. The first Mars missions (e.g., Mariner 9 [7]) reached the science orbit with an all-propulsive strategy corresponding to total $\Delta$Vs of about 1.6 km/s. In more recent missions, such as Mars Global Surveyor (MGS) [8] and Mars Reconnaissance Orbiter (MRO) [9], a MOI of 1 km/s was supplemented by aerobraking campaigns of several months to circularize the orbit and reduce the orbital period.

According to a recent NASA's study [1], aerocapture is ready to be employed for science missions at Venus, Mars and Titan and it may enable missions to Uranus and Neptune after further studies and developments. In the case of Mars, a high technological readiness has been reached through many missions performing atmospheric entry. However, despite all its potential benefits, aerocapture has never been implemented because of the uncertainties in Mars atmospheric density and its variations as well as navigation errors [10]. Nevertheless, demonstrations relevant to aerocapture have already been accomplished. For instance, in the '60s, the hypersonic guidance and control necessary for a skip entry into Earth's atmosphere was demonstrated by the Apollo 6 vehicle and, in the same years, the Soviet Zond spacecraft performed a successful skip entry from lunar return. In 2014, the Chang'e 5-T1 mission by the Chinese space agency performed a similar aeroassisted manoeuvre. In August 2012, NASA's Mars Science Laboratory (MSL) demonstrated the capability of accurate soft-landing on the Martian surface by using autonomous hypersonic guidance in the atmosphere. Subsequent studies about the relative difficulty of skip entry, hypersonic manoeuvring and aerocapture have led to conclude that both skip entry and hypersonic manoeuvring to precision landing are more challenging than aerocapture because of the tighter tolerances on the vehicle's state [1].

Aerocapture was adopted in the early phases of the design of 2001's Mars Odyssey mission [11], but then replaced with an aerobraking strategy when issues related to the packaging of the aeroshell and the mass budget arose. Since then, the possibility to employ aerocapture for small satellites has been investigated thoroughly. A recent study has demonstrated that aerocapture for SmallSats, i.e., satellites with a mass lower than 180 kg, could increase the delivered mass to Mars, Venus and other destinations [12]. In particular, drag-modulation aerocapture turned out to be an interesting option for SmallSat missions at Mars, since the associated low heat rates allow to use large and lightweight inflatable drag devices [13]. Falcone et al. [14] have shown that single-event drag-modulation aerocapture could be a good way to improve mission flexibility for SmallSat missions at Mars by decreasing the orbit insertion system mass fraction by 30% or more with respect to propulsive options.

The focus of this work is on the feasibility of aerocapture at Mars with an innovative deployable drag device. The device is a deployable heat shield (DHS) designed for the Small Mission to MarS (SMS) project [15,16] on the basis of the Italian ReEntry NacellE (IRENE) technology developed for Earth re-entry missions [17,18]. The DHS was designed to provide deceleration and thermal protection during the entry phase, but could be used also for orbit insertion by means of aerocapture.

The analysis here presented is conducted by a purely dynamical point of view, coming alongside with previous studies about the thermal and structural design of the DHS. The analysis focuses on the Mars arrival phase and takes into account the main uncertainties affecting the problem. We first present a parametric bi-dimensional analysis, covering a wide range of different possible mission scenarios. By construction, aerocapture is a two-dimensional maneuver, so a 2D analysis is useful to draw general considerations about its feasibility. This is followed by a three-dimensional study for a specific SMS mission scenario. Both investigations are carried out by varying the density profile and the ballistic coefficient of the capsule, in this way simulating atmospheric and navigation uncertainties. Assessing the effect of the uncertainties on the feasibility of the aerocapture maneuver is the purpose of the study.

Section 2 describes the approach, the dynamical model and the adopted atmospheric density profile and reviews the main Mars atmospheric density models. Section 3 contains an overview of the SMS mission, its objectives and the system architecture with particular focus on the DHS. Section 4 presents a description of the parametric two-dimensional analysis and its results. Finally, in Section 5 the results of the application to a specific mission scenario are illustrated, while Section 6 draws the conclusions.

## II. Aerocapture analysis approach

This work focuses on the final phase of an interplanetary trajectory from Earth to Mars modelled with patched conics (Fig. 1) and, more specifically, it deals with the analysis of the aerocapture manoeuvre. For this reason, the simulations presented in this paper start at the surface of Mars' Sphere of Influence (SOI).

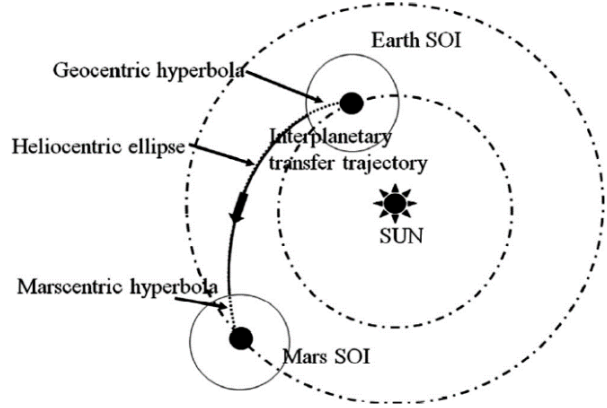

**Fig. 1    Patched conics approximation for an Earth-to-Mars direct interplanetary transfer [19].**

In the patched conics approximation, after entering Mars' SOI, the spacecraft is assumed to be only subject to the gravitational attraction of Mars, i.e., its trajectory is modelled as a gravitational two-body orbit and the solution is a Mars-centric hyperbola. In this work, the arrival hyperbola is propagated up to the atmospheric interface (AI), for which an altitude of 125 km above Mars' surface is assumed (a value of common use in the literature) (Fig. 2). Below this limit, the physical model is the drag-perturbed two-body problem,

$$\ddot{\boldsymbol{r}} + \frac{\mu}{r^3}\boldsymbol{r} = \boldsymbol{a}_D \ , \tag{1}$$

where $\boldsymbol{r}$ is the position vector of the spacecraft with respect to the center of Mars, $\mu$ is the standard gravitational parameter of Mars and $\boldsymbol{a}_D$ is the acceleration caused by atmospheric drag. The drag perturbation is expressed as follows

$$\boldsymbol{a}_D = -\frac{1}{2}\frac{\rho}{\beta}v\boldsymbol{v} \ , \tag{2}$$

$\rho$ being the atmospheric density, a function of altitude, $\beta$ the ballistic coefficient of the probe, $\boldsymbol{v}$ the relative velocity of the spacecraft with respect to the atmosphere and $v$ its magnitude. The ballistic coefficient $\beta$ is defined as

$$\beta = \frac{m}{C_D A}, \tag{3}$$

where $C_D$ is the drag coefficient, $m$ is the mass and $A$ is the cross section of the spacecraft. The last term can be expressed as $A = A_{max} \cos\theta$, with $A_{max}$ the maximum drag cross-section representing the nominal configuration and $\theta$ the angle of misalignment with respect to such configuration.

Other perturbations act on the trajectory of the spacecraft, such as the inhomogeneous mass distribution within the planet and the gravitational attraction of other bodies in the Solar System. Regarding the former, this study considers the acceleration $a_{J2}$ caused by the second-degree term $J_2$ of the mass distribution of Mars, whereas the third-body perturbation is limited to the contribution $a_{Sun}$ due to the Sun. The expressions for the respective additions to Eq. (1) are

$$\boldsymbol{a}_{J2} = -\frac{1}{2}\frac{\mu J_2 R^2}{r^5}\left[1 - 3\sin^2(\phi)\right]\boldsymbol{r}, \tag{4}$$

and

$$\boldsymbol{a}_{Sun} = \mu_{Sun}\left(\frac{\boldsymbol{R}_{Sun} - \boldsymbol{r}}{\|\boldsymbol{R}_{Sun} - \boldsymbol{r}\|^3} - \frac{\boldsymbol{R}_{Sun}}{\|\boldsymbol{R}_{Sun}\|^3}\right), \tag{5}$$

[20]. Here $R$ is the Mars' radius, $\Phi$ is the geocentric latitude of the spacecraft with respect to Mars, $\mu_{Sun}$ is the standard gravitational parameter of the Sun and $R_{Sun}$ is the position vector of the Sun with respect to Mars' center.

The solutions of Eq. (1) and those obtained with the addition of the terms given by Eqs. (4) and (5) are approximated numerically by using a Runge-Kutta scheme of $4^{th}$-$5^{th}$ order. The model described by Eq. (1) is adopted in the 2D analysis. The extended model, i.e., that including $J_2$ and third-body effects, is used to assess the outcome of the additional perturbations on the trajectories corresponding to the specific mission scenario considered in the 3D study. At atmospheric exit (AE), the orbital parameters are computed to verify if the capture has been achieved, i.e., if the resulting orbit is elliptical.

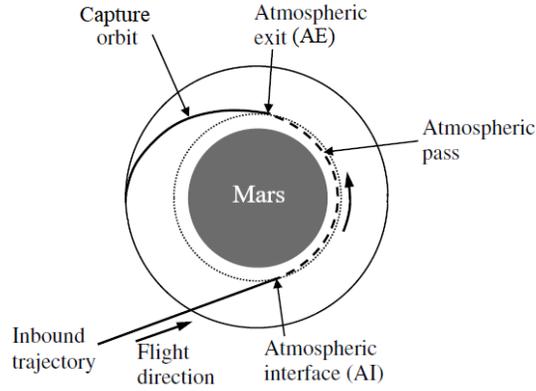

Fig. 2  Schematic illustration of the aerocapture maneuver.

In the bi-dimensional case, a parametric study is carried out by changing the magnitude $v_\infty$ of the hyperbolic excess velocity $\mathbf{v}_\infty$ at Mars arrival and the periapsis altitude $h_\pi$. In this way, a wide range of trajectories can be examined, and general conclusions can be drawn. All the parameters are expressed in the Perifocal Reference Frame centered at Mars (see Fig. 3): the $\hat{p}$ unit vector points to pericenter, $\hat{w}$ is aligned with the orbital angular momentum and $\hat{q}$ completes the right-handed triad.

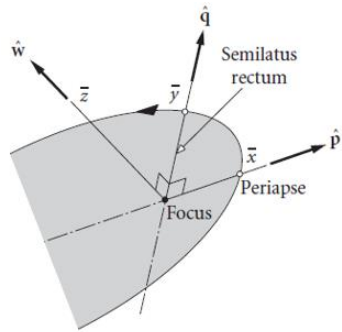

Fig. 3  The Perifocal Reference Frame.

The three-dimensional analysis, instead, is carried out for a specific SMS mission scenario (see Section 3). The state vector at Mars' SOI corresponding to an arrival option in September 2025 (and Earth departure 11 months earlier) is used as the initial condition. Since the initial hyperbolic trajectory targets the centre of Mars, an impulsive retargeting manoeuvre is applied at the surface of the SOI. The spacecraft is made to pass through the atmosphere at altitudes that could be suitable for aerocapture while avoiding impact with the surface. All the computations are performed in the Mars Mean Equator of Date frame based on IAU 2000 Mars Constants (MMEIAU2000) (Fig. 4) [21,22]. The origin of this frame is the centre of Mars and its $xy$-plane is the Mars mean equatorial plane. The $z$-axis is in the direction of the Mars' rotation axis, whose orientation is given as equatorial coordinates in the International Celestial Reference Frame, whereas the *x*-axis is defined by the intersection between Mars' equator and Earth's mean equator of epoch J2000. The *y*-axis completes the right-handed system.

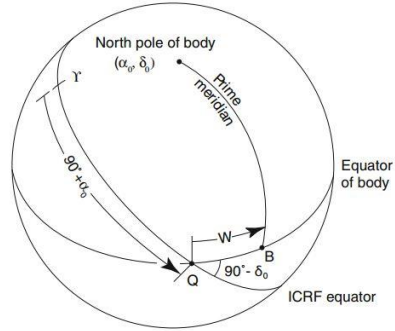

**Fig. 4    Reference system used to define the orientation of Mars' rotation axis and the MMEIAU2000 frame.**

[22]

Regarding the atmospheric modeling, the density profile chosen for this study is the Mars Global Reference Atmospheric Model (Mars-GRAM) [23,24], which provides the density for any location (altitude, latitude and longitude) and time (down to daily variations). It is an engineering-oriented empirical model of the Mars atmosphere and it was built with the parametrization of the data collected by Mariner and Viking and the results of NASA Ames Mars General Circulation Model (MGCM) for altitudes between 0 and 80 km, and the University of Arizona Mars Thermospheric General Circulation Model (MTGCM) above 80 km [25,26]. Mars-GRAM computes atmospheric parameters (temperature, pressure, density), surface physical data, electromagnetic fluxes, atmospheric heating rates, as well as thermodynamic properties of $CO_2$. The 2001 version of Mars-GRAM is the core component of the Mars Environment Multi-Model (MEMM) [27], an engineering tool developed at the Polytechnic University of Catalonia, which merges existing models of the Mars environment. MEMM outputs the several atmospheric parameters in a unified way (see also Appendix A).

The adoption of Mars-GRAM follows a detailed comparison with other state-of-the-art density models based on atmospheric data obtained from previous Mars missions. In order to compare the various models and evaluate the capability of Mars-GRAM to simulate all the possible atmospheric conditions, two density profiles have been extracted from MEMM, respectively representing a very high density and a very low density case and corresponding to one hot and one cold case scenario under clear sky conditions, i.e., away of dust storms. The hot case is a summer (Solar Longitude $L_s$ = 270°) midday profile under high solar activity (Solar F10.7 Flux = 300 SFU). The cold case, instead, is a winter ($L_s$ = 90°) nighttime profile at low solar activity (Solar F10.7 Flux = 60 SFU).

A brief review of the models considered in the comparison (Fig. 5) is given below. First of all, we have considered the atmospheric density data collected in 1976 by the Viking 1 lander during its atmospheric descent and the profile based on the relevant observations collected continuously between 150 and 10 km of altitude by the Mars Pathfinder lander in 1997 [28]. Then, the density profile provided by COSPAR Mars Reference Atmosphere, as reported in [29], has been considered. ESA's 2003 Mars Express orbiter observed 616 stellar occultations, deducing the amount of $CO_2$, the main constituent of the Martian atmosphere, between the instrument and the star, with its ultraviolet spectrometer SPICAM from January 14, 2004 (Mars Solar Longitude, $L_s$ = 332.8°) to April 11, 2006 ($L_s$ = 37.6°) [30]. [30] reports three density profiles corresponding to different epochs and positions, but only the most extreme profile has been used for the comparison.

The densities predicted by the several models allow to define a range of values at each altitude. According to Fig. 5, the cold case Mars-GRAM density is always the lowest and for this reason here it has been adopted as the lower bound of the density ranges for all the altitudes of interest. The hot case Mars-GRAM density is not always the highest, instead. First of all, there are altitudes at which the Viking density is higher than the Mars-GRAM hot case, but the differences are small, and Mars-GRAM includes Viking observations as well. There is also a window of altitudes, between 75 and 110 km, in which the SPICAM density profile overtakes the hot case Mars-GRAM profile. A possible explanation for this phenomenon is the variation of the dust content of the lower atmosphere, which causes most of the density seasonal variations. Indeed, other profiles from SPICAM obtained for different Solar Longitudes show lower density values and never exceed Mars-GRAM's hot case [30]. For these reasons and for the sake of simplicity, Mars-GRAM has been adopted also to represent the upper bound of the density range.

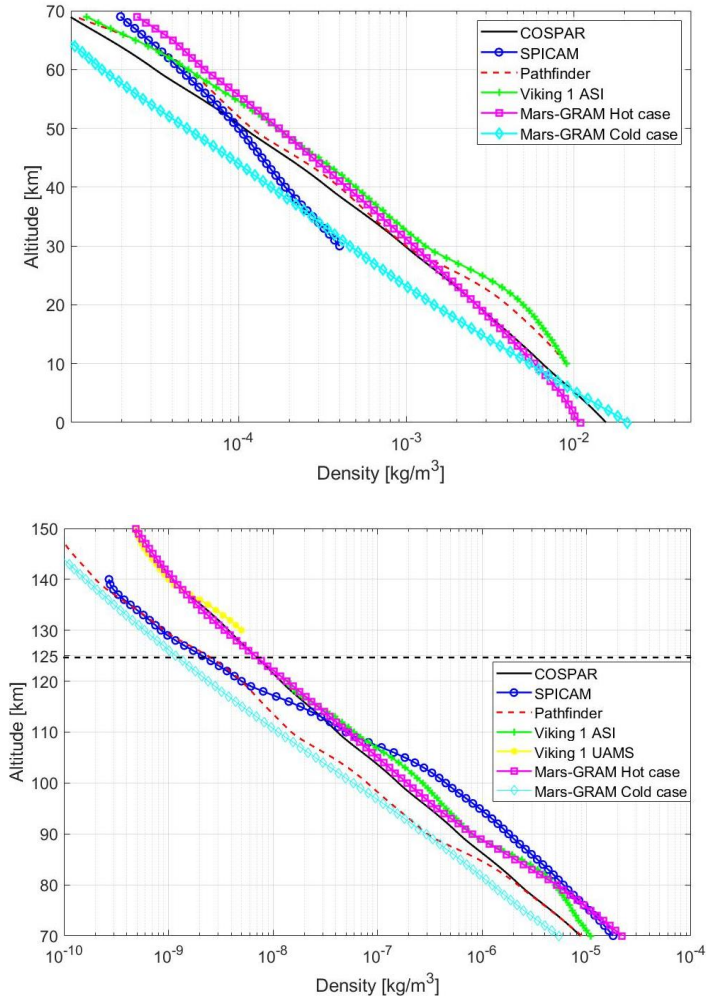

**Fig. 5  Comparison among Mars atmospheric density profiles as predicted by various models for the altitude range 0 to 70 km (top) and 70 to 150 km (bottom). The dashed black line at 125 km indicates the assumed upper limit of the atmosphere.** [31]

The nominal density profile selected for the analysis is shown in Fig. 6. It has been provided by MEMM for August 29th 2025, the date of the entry of SMS into Mars' SOI, in clear sky conditions at noon.

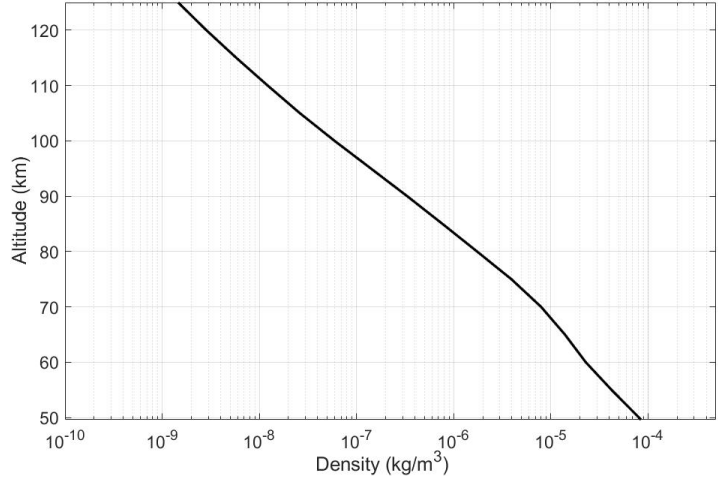

**Fig. 6  Nominal Mars-GRAM atmospheric density profile for the altitudes of interest (results output by MEMM).**

### III.  The SMS mission

SMS was initially conceived as a landing mission. Its development started within ESA's General Support Technology Programme (GSTP). The main objective of SMS was a technology demonstration mission to Mars characterized by small size and low cost. Designed to fit inside a VEGA rocket, it featured an innovative DHS for system deceleration and thermal protection during entry, descent and landing (EDL). Most of the information in this section is extracted from [15,16,31].

The SMS spacecraft presents a modular architecture consisting of the capsule, containing the lander module, stowed inside the DHS during launch and interplanetary cruise, the cruise stage (CS) providing power, communications and propulsion support during interplanetary journey, and a kick stage for orbit raising and interplanetary injection (Fig. 7, left). The DHS is deployed by an umbrella-like mechanism just before AI (Fig. 7, center and right). In this way, the mass/volume ratio at launch increases, widening the choice of possible launchers and, in particular, leading to the possibility of using a small launcher, such as VEGA. The wet system mass is 300 kg, whereas the mass left after jettisoning the CS at Mars approach is 150 kg. In the early SMS design, launch and arrival options were determined in a Sun-spacecraft two-body model following a direct transfer from Earth to Mars. The chosen solution was the minimum-cost option within the 2024 launch window. It is a type II transfer, i.e., the transfer angle is larger than 180°, and corresponds to departing on October 2$^{nd}$ 2024 and arriving on September 1$^{st}$ 2025. This solution has an Earth $C_3$ (launch energy) of 11.316 km$^2$/s$^2$ and a Mars $v_\infty$ of 2.455 km/s. The only deterministic manoeuvre along the trajectory was the targeting manoeuvre at the Mars' SOI characterized by a magnitude between 33 and 50 m/s.

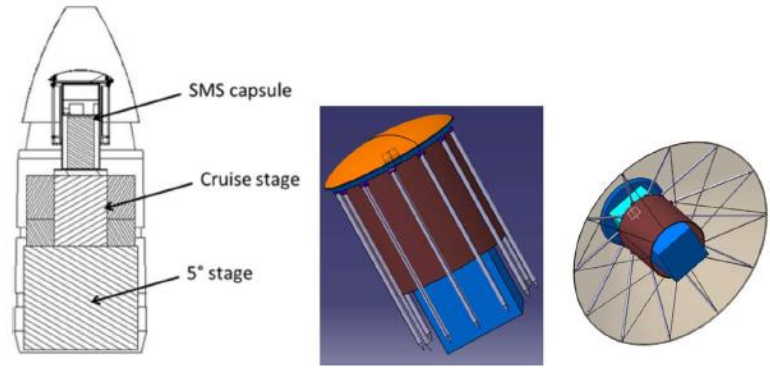

**Fig. 7** Left: SMS capsule (Lander + DHS), CS and kick stage after assembly and packing in the payload fairing of VEGA. Center and right: the lander inside the DHS in folded and unfolded configuration, respectively.

In the successive phases of the mission development, several innovations have been introduced, including the possibility of using the DHS for Mars orbit insertion through an aerocapture manoeuvre, eventually followed by an aerobraking campaign to lower the eccentricity of the elliptical orbit around Mars. The significant propellant savings resulting from this new development would increase the payload mass fraction and allow to transfer to Mars additional scientific payloads, including CubeSats to be released before atmospheric entry.

The DHS, whose characteristics are described in [15], provides both thermal protection and deceleration through the atmosphere and has been designed on the basis of the IRENE capsule concept [17]. IRENE, which can be seen as a first version of the DHS, was conceived for terrestrial applications in the context of research and industrial projects in the field of suborbital re-entry technology [32] and it is currently in flight test phase.

In all past Mars missions, the design of the entry capsule always included a fixed forebody heat shield, which protected it from the high aerodynamic heating of atmospheric entry. The advantage of using a deployable umbrella-like device is the small attainable ballistic coefficient. The ballistic coefficient achievable by SMS with the DHS can be as small as 20 kg/m$^2$, i.e., less than one third the value for any previous Mars mission [33] (Fig. 8 and Table 1), allowing higher decelerations even in the upper region of Mars atmosphere and a substantial reduction of the aerodynamic and aero-thermo-dynamic loads during the hypersonic entry flight path. The umbrella-like concept in the DHS is similar to the ADEPT technology developed by NASA [34,35].

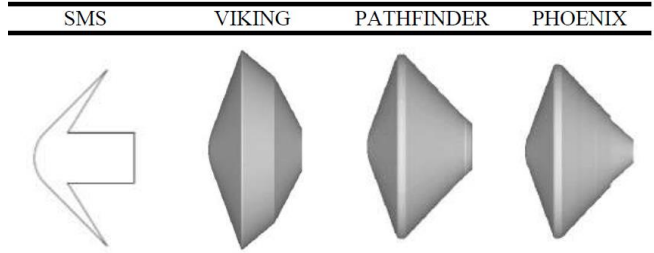

**Fig. 8   Comparison of the shape of DHS of SMS with previous Mars entry systems.**

**Table 1 Comparison of the characteristics of DHS of SMS with previous Mars entry systems.**

| Parameter | SMS | Viking | Pathfinder | Phoenix |
|---|---|---|---|---|
| Relative entry velocity, km/s | 5.5 | 4.5 | 7.6 | 5.9 |
| Relative entry flight path angle, ° | -13 | -17.6 | -13.8 | -13.2 |
| Ballistic coefficient, kg/m$^2$ | 21 | 64 | 62 | 65 |
| Entry altitude, km | 125 | 82 | 125 | 125 |

## IV.  Parametric bi-dimensional analysis

The parametric analysis is realized by varying the magnitude $v_\infty$ of the hyperbolic excess velocity $\boldsymbol{v}_\infty$ and the periapsis altitude $h_\pi$. These two quantities uniquely define the shape of the arrival hyperbola. Due to the symmetry of the problem, the study is limited to zero-inclination Mars arrival hyperbolic orbits (i.e., in Mars equatorial plane). The parameters used in these simulations are those of the SMS capsule. In particular the ballistic coefficient $\beta$ is set at 21.23 kg/m$^2$ and $C_D = 1$ [36]. The parametric analysis is performed for the $v_\infty$ and $h_\pi$ ranges of Table 2 and using the nominal Mars-GRAM atmospheric density profile extracted from MEMM. Then, the uncertainties in the atmospheric density and ballistic coefficient are included (see Table 3). We simulate the density uncertainty by increasing and decreasing the density values by a percentage amount. The Mars-GRAM density profile extracted from MEMM and the density profiles obtained considering uncertainties up to 70% of the nominal value are shown in Fig. 9. The uncertainty of 6.38 kg/m$^2$ in the ballistic coefficient reported in Table 3 is representative of variations of about 30% in the aerodynamic drag coefficient and about 10° in $\theta$, including attitude control errors and errors in the shield opening mechanism.

**Table 2 $v_\infty$ and $h_\pi$ ranges and resolutions used in the 2D parametric analysis.**

| Parameter | Range | Increment |
|---|---|---|
| Hyperbolic excess velocity, $v_\infty$, km/s | 2 to 4 | 0.1 |
| Periapsis altitude, $h_\pi$, km | 50 to 110 | 0.1 |

**Table 3 Uncertainties in the atmospheric density and ballistic coefficient.**

| Parameter | Uncertainty |
|---|---|
| Atmospheric density | ±35%, ±50% and ±70% with respect to the nominal density profile |
| Ballistic coefficient, kg/m$^2$ | ±6.38 |

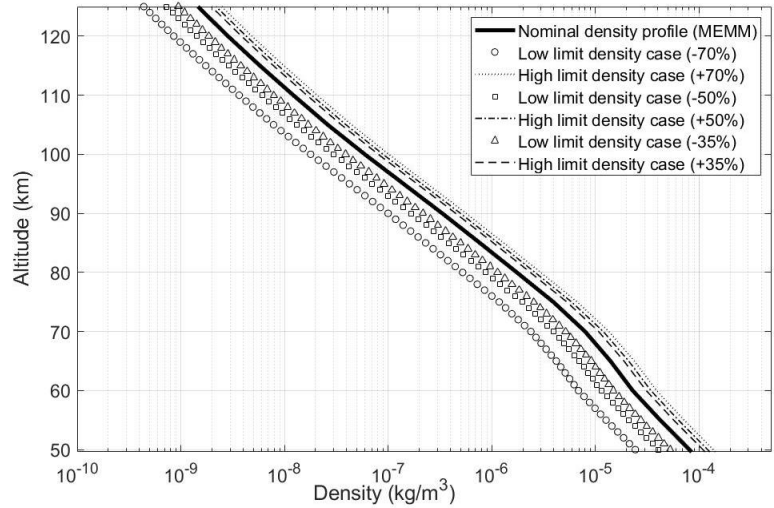

**Fig. 9   Atmospheric density profiles for the range of altitudes of interest in this study.**

Each $v_\infty$-$h_\pi$ pair in the selected ranges corresponds to a different hyperbolic arrival trajectory. A first simulation was carried out for nominal atmospheric conditions, i.e., without any uncertainty, and the results show a range of periapsis altitudes producing aerocapture for each arrival velocity in the given interval (Fig. 10). It can be noted that as $v_\infty$ increases, the intervals of useful $h_\pi$ shrink and the centers of the intervals move towards lower altitudes. The periapsis altitudes reported in Fig. 10 are those of the ideal Keplerian trajectories, whereas the actual periapsis altitudes are slightly lower. The gap between the Keplerian and the real $h_\pi$ varies from more than 0.6 km for altitudes around 75 km to a few tens of meters for altitudes around 95 km.

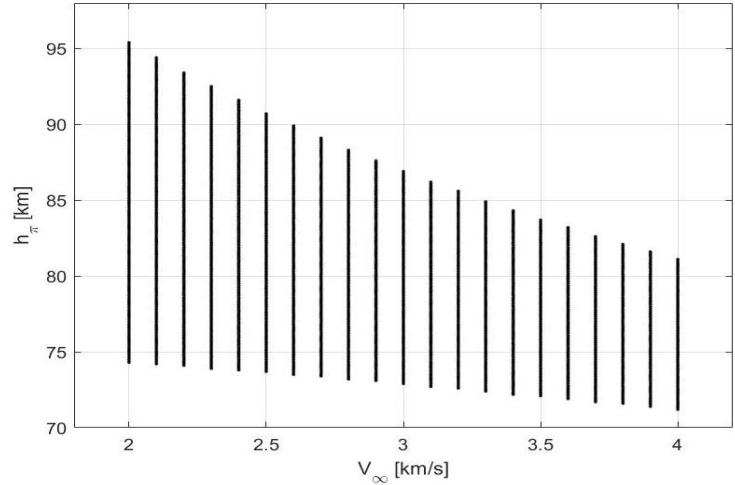

Fig. 10  Map of $v_\infty$-$h_\pi$ pairs leading to aerocapture in the nominal scenario.

Figure 11 illustrates the results obtained by considering different levels of atmospheric density uncertainty and keeping the ballistic coefficient at its nominal value. For each uncertainty range, only the two extreme density conditions (vertex analysis) are simulated and only the common solutions are included. The number of successful pairs decreases with increasing level of uncertainty. For uncertainty values higher than 35% there are arrival velocity intervals which do not lead to capture. However, we note that even with 70% atmospheric density uncertainty it is still possible to find viable solutions for aerocapture when the arrival velocity is between 2 and 2.5 km/s and the periapsis is above 80 km altitude.

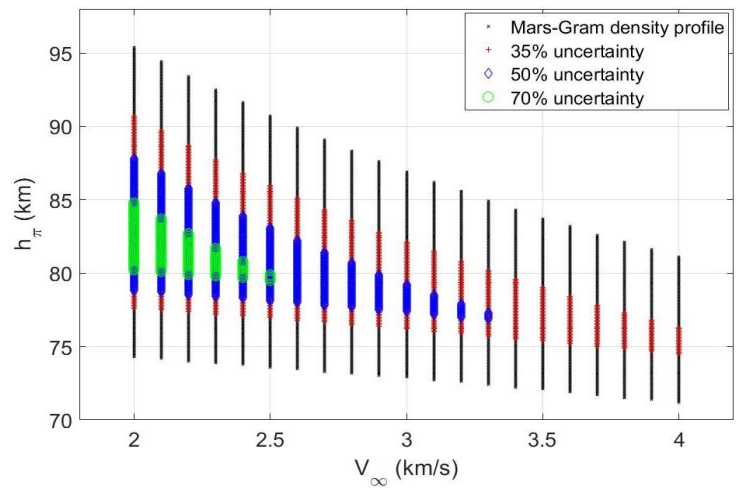

Fig. 11  Map of $v_\infty$-$h_\pi$ pairs leading to aerocapture for the nominal scenario and its modification with increasing atmospheric density uncertainty.

Figure 12 shows the effect of the ballistic coefficient uncertainty on the solutions computed with the nominal atmospheric density profile. Also here only common solutions are considered. It can be noticed that, in general, the periapsis altitude range leading to aerocapture is reduced, but an interval for this parameter can still be found for each arrival velocity.

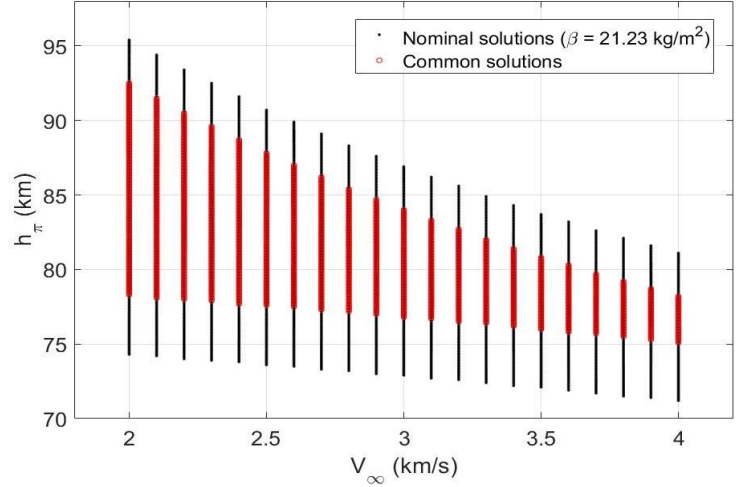

**Fig. 12   Map of $v_\infty$-$h_\pi$ pairs leading to aerocapture for the nominal scenario and its modification when the uncertainty in the ballistic coefficient is introduced.**

Finally, we analyse the simultaneous effect of ballistic coefficient and atmospheric density uncertainties. In this case, aerocapture is achieved only for atmospheric uncertainties within ±35% (Fig. 13), but they are limited to Mars arrival velocities up to 2.8 km/s and periapsis altitudes higher than 80 km. It can be noticed the significant narrowing of the altitude ranges producing aerocapture.

In summary, we have found combinations of parameters which guarantee the success of the aerocapture manoeuvre regardless of the uncertainties, when these are within particular ranges, and this represents a general result for the bi-dimensional analysis.

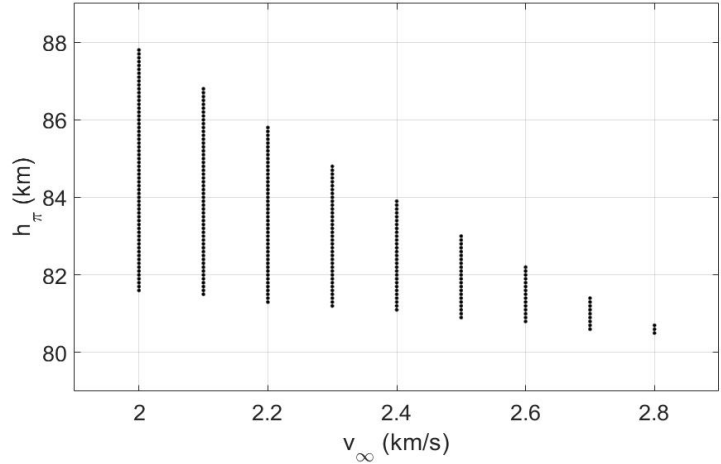

**Fig. 13 Map of $v_\infty$-$h_\pi$ pairs that lead to captured orbits when both the ballistic coefficient uncertainty and a 35% uncertainty in the atmospheric density are included.**

## V. Application to a SMS mission scenario

The initial conditions for the study are the orbital parameters of the spacecraft when it enters the SOI according to the nominal SMS mission [15] (Table 4). They correspond to a $v_\infty$ of 2.455 km/s.

**Table 4 Orbital parameters of the hyperbolic trajectory at Mars' SOI.**

| Parameter | Value |
| --- | --- |
| Epoch (UTC) | 29/8/2025   06:42:11 |
| Semi-major axis, km | -7287.768 |
| Eccentricity | 1.0000274 |
| Inclination, (°) | 98.8 |
| Right Ascension of the Ascending Node, (°) | 122.5 |
| Argument of Pericenter, (°) | 263.7 |
| True Anomaly, (°) | 180.4 |

The analysis consists in applying a velocity variation at the SOI of Mars targeting an atmospheric passage at altitudes suitable for aerocapture, in agreement with the 2D parametric analysis. A loop is considered on the three parameters characterizing the manoeuvre, i.e., its magnitude ΔV and its orientation as expressed by two angles, λ and δ (see Table 5), respectively representing longitude and latitude in a reference frame with origin at the spacecraft centre of mass and parallel to MMEIAU2000. The range of variation of ΔV is consistent with the magnitude of the nominal targeting manoeuvre of SMS and with the propulsive limitations of a platform of its category.

**Table 5 The parameters of the targeting manoeuvre.**

| Parameter | Range | Increment |
|---|---|---|
| Magnitude, $\Delta V$, m/s | 30 to 40 | 0.1 |
| Longitude, $\lambda$, (°) | 0 to 360 | 1 |
| Latitude, $\delta$, (°) | -90 to 90 | 1 |

The values reported in Table 5 are intended for a first-look analysis. In a second phase, $\Delta V$ is kept constant and equal to the value that provides continuity in the two angles when these are varied over ranges of interest at increments of 0.1°. Since this value is representative of the angular accuracy attainable in the execution of the manoeuvre, checking the continuity against these smaller angular increments is crucial to ensure the success of the aerocapture even in presence of manoeuvring errors.

The analysis starts from the same nominal atmospheric density profile and ballistic coefficient and the same uncertainties as in Table 3. Applying the manoeuvres of Table 5 to the nominal conditions yields the 3D map of successful captures displayed in Figure 14. It can be noted that aerocapture is successful only when the magnitude of the impulse is 33.4 m/s or larger. Solutions become more and more scattered as $\Delta V$ increases. Thus, the manoeuvres with $\Delta V$ of 33.4 m/s are the most interesting because they give rise to successful captures over continuous ranges of the two angles. The 2D map of the successful manoeuvres for the selected $\Delta V$ is shown in Fig. 15.

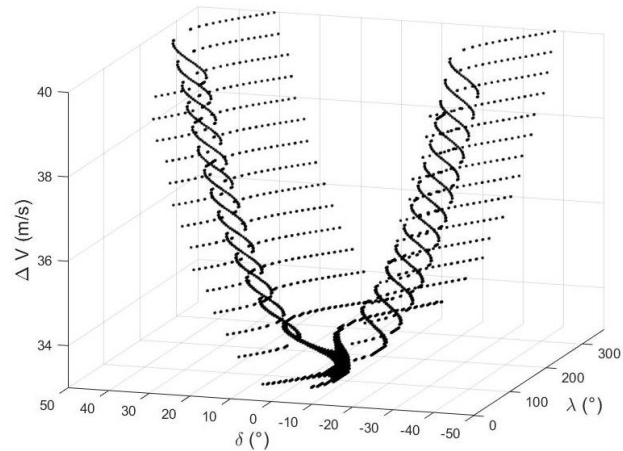

**Fig. 14 3D map of the manoeuvres leading to aerocapture when considering the ranges and the increments of the manoeuvre's parameters reported in Table 5.**

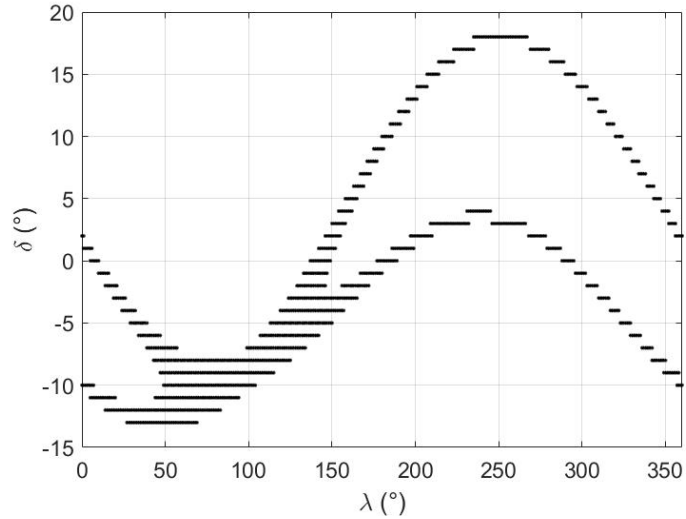

**Fig. 15  2D map of the successful orientations of the targeting manoeuvre when the impulse has a magnitude ΔV of 33.4 m/s.**

The solutions are scattered everywhere except in the region corresponding to the range 20° to 180° in λ and -15° to 0° in δ. In this subset, the continuity is checked against smaller variations in both angles by applying a discretization of 0.1°, representative of the angular accuracy of the manoeuvre. Figure 16 shows that the continuity of the successful manoeuvre orientations persists at this higher angular resolution.

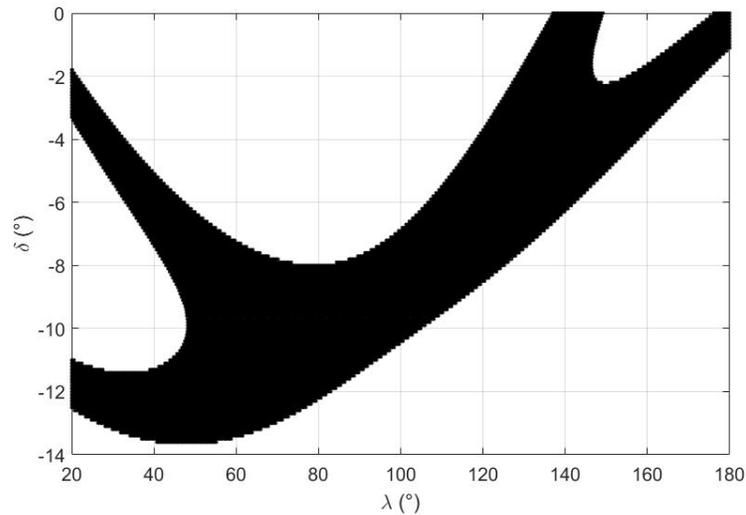

**Fig. 16  2D map of the successful orientations of the targeting manoeuvre when the impulse has a magnitude ΔV of 33.4 m/s and the angular resolution is 0.1°.**

Figure 17 illustrates the periapsis altitudes $h_\pi$, i.e., the minimum altitudes reached in atmosphere, of the hyperbolic trajectories obtained after application of the velocity variation at Mars' SOI for the ranges of the parameters that ensure continuity in the higher angular resolution exploration. The outcome is in good agreement with the results obtained in the 2D analysis for the same nominal scenario (Fig. 10).

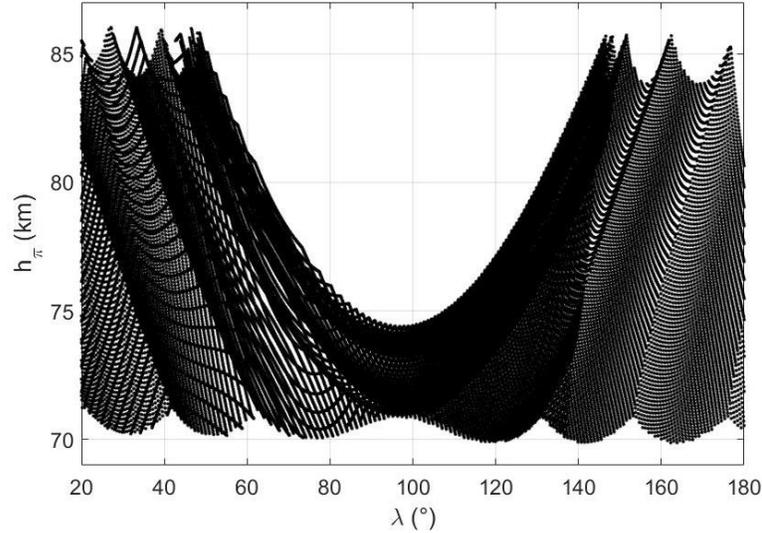

**Fig. 17 Map of periapsis altitudes $h_\pi$ of the hyperbolic trajectories suitable for aerocapture after the application of the manoeuvre at Mars' SOI.**

Then, keeping fixed the ballistic coefficient at the nominal value, the influence of the different levels of atmospheric density uncertainty on the distribution of solutions in the λ-δ map of Fig. 16 has been investigated. Like for the parametric analysis, for each uncertainty level, we accept only the solutions in common to the two vertex density conditions. The results are shown in Fig. 18. The ±70% uncertainty case gives no solutions. The reason is that at -70% uncertainty, the atmosphere is very thin and does not cause a sufficient deceleration. Also, in the 2D analysis the number of solutions obtained for this level of uncertainty is small (see Fig. 11). Figure 18 has been obtained with increasing atmospheric density uncertainty (nominal, 35% and 50%): the number of successful combinations of parameters decreases as indicated by the shrinking of the corresponding areas in the map. The initial continuity is lost.

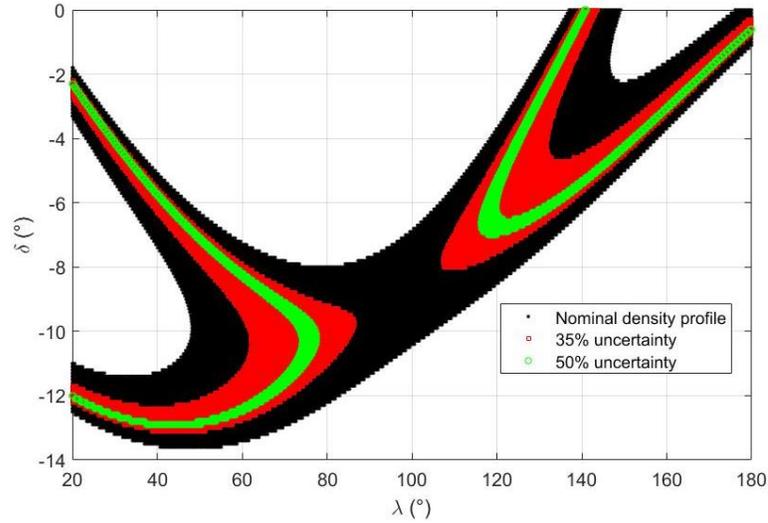

**Fig. 18 2D map of the successful orientations of the targeting manoeuvre with increasing atmospheric density uncertainty (nominal, 35% and 50%).**

Since at uncertainties of -50% and -70% the number of successful parameter combinations is very low, we have limited the analysis to uncertainties between -35% and +70%. The corresponding map is displayed in Fig. 19. The simulations carried out for the two extreme density cases in this uncertainty interval lead to very different elliptical orbits in terms of eccentricity $e$ and apocenter altitudes $h_\alpha$, as shown in Figs. 20 and 21. When assuming +70% of density uncertainty, the resulting orbits have $e$ lower than 0.3 and $h_\alpha$ limited to 2500 km, while, for the -35% density uncertainty case, the eccentricities are higher than 0.9.

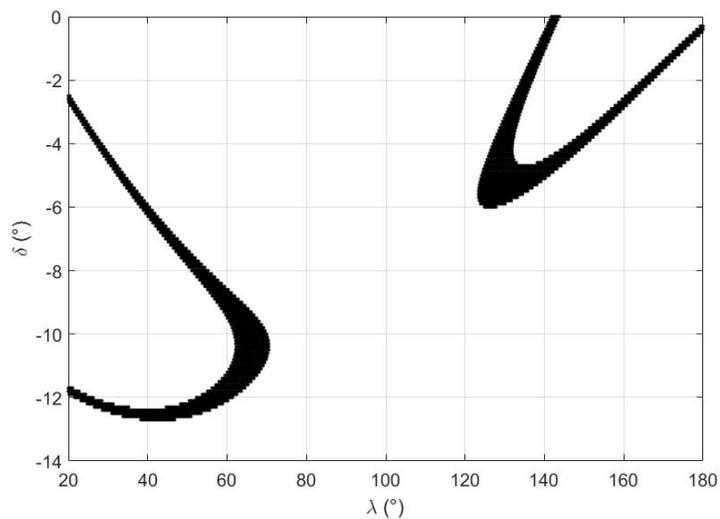

**Fig. 19 2D map of the successful orientations of the targeting manoeuvre for atmospheric density uncertainties between -35% and +70%.**

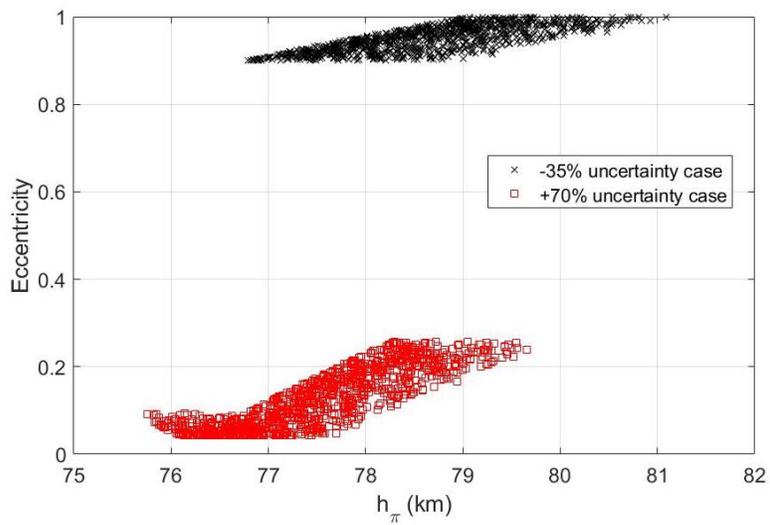

**Fig. 20   Eccentricities of the captured orbits at -35% (black crosses) and +70% (red squares) density uncertainty, respectively, versus the minimum heights reached in the atmosphere.**

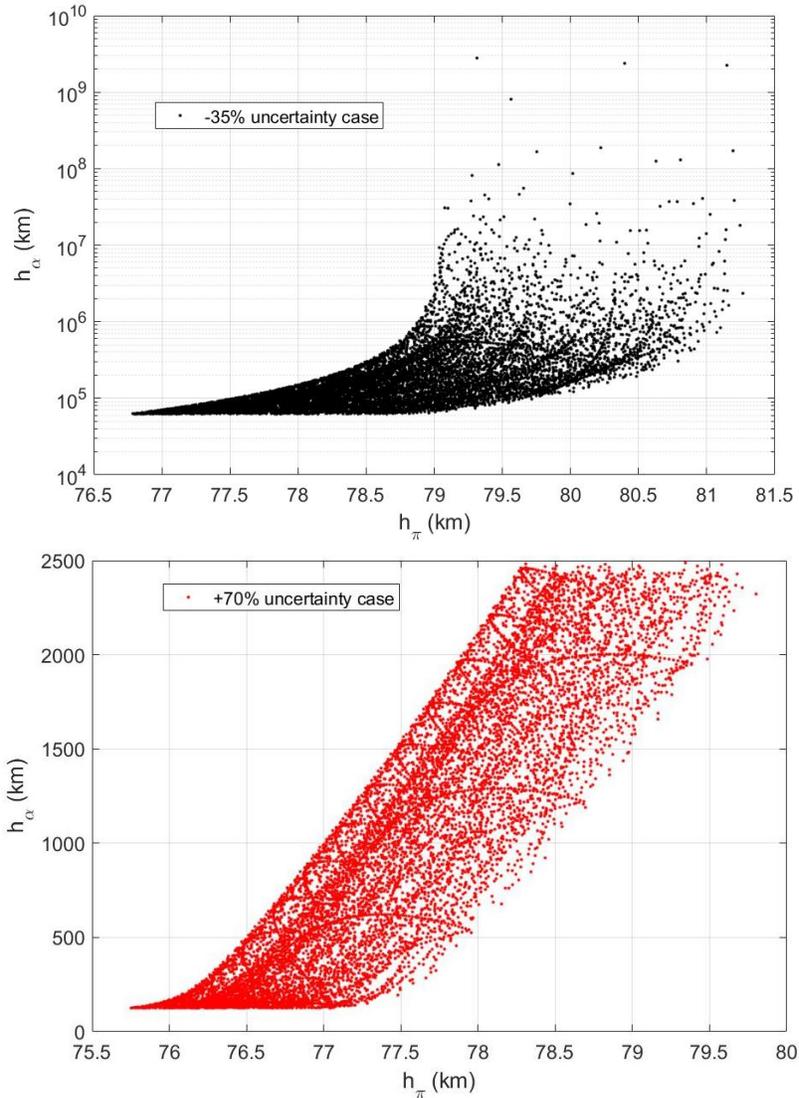

**Fig. 21  Apocenter altitudes of the captured orbits at -35% and +70% density uncertainty, respectively in the top and bottom plots, versus the minimum heights reached in the atmosphere.**

Finally, Fig. 22 illustrates the effect of the uncertainty in the ballistic coefficient. When the ballistic coefficient is lower or higher than the nominal value, the areas of successful parameter combinations become wider and tend to shift slightly. This is very interesting since a changing ballistic coefficient can be achieved by increasing or reducing the drag cross section through in-flight modulation of the aperture of the DHS.

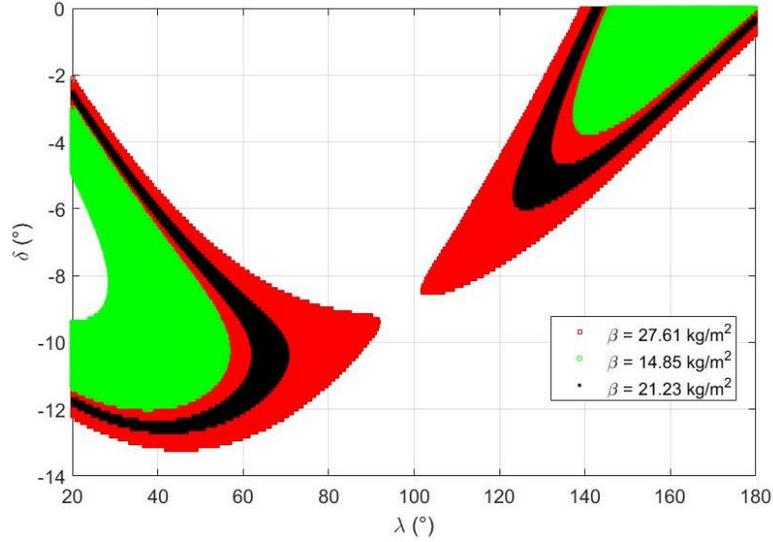

**Fig. 22 Effect of the ballistic coefficient uncertainty on the combinations of angles that correspond to captured orbits for density uncertainties between -35% and +70%.**

For the sake of completeness, an example of circularization manoeuvre is presented in the following to derive the order of magnitude of the ΔV budget of the entire mission for a case of possible scientific interest. Considering the nominal scenario, applying for instance the manoeuvre characterized by λ = 100° and δ = -10.2°, the spacecraft exits the atmosphere in an orbit with $h_α$ of 692.8 km and $e$ of 0.0871. The periapsis raising manoeuvre needed to circularize this orbit has a size of 144.2 m/s, which brings the total ΔV budget to 177.6 m/s.

## VI. Conclusion

This contribution is an assessment of the aerocapture at Mars with a deployable drag device considering the main uncertainties affecting the manoeuvre. A parametric bi-dimensional analysis carried out over a wide range of conditions has been supplemented and completed by an application to a real mission scenario in which the targeting manoeuvre leading to successful aerocapture is studied. In both analyses, the atmospheric and navigation uncertainties have been simulated through variations in the atmospheric density profile and the ballistic coefficient of the capsule, respectively. In this way, their effect on the feasibility of the aerocapture manoeuvre has been assessed.

The 2D analysis has pointed out the significant impact of the atmospheric density uncertainty, which considerably reduces the solution space when variations of more than 50% of the nominal density profile are assumed. Adding the uncertainty on the ballistic coefficient leads to captured orbits only for density variations up to 35% and for low Mars arrival velocities (below 2.8 km/s).

The 3D analysis conducted on the SMS mission has shown that captured orbits can be obtained over a wide, continuous parameter subset for a specific value of the magnitude ΔV of the targeting manoeuvre at the surface of Mars' SOI. It also confirms the strong impact of atmospheric density and ballistic coefficient uncertainties already shown by the 2D analysis. Solution continuity is an important element, considering the possible errors in the execution of the targeting manoeuvre. In any case, the results of the simulations prove that aerocapture is viable for density errors in the range -35% to +70% even in the presence of ballistic coefficient uncertainties. The role played by the shield in ensuring this result means that the deployable drag device here considered is suitable to perform aerocapture.

Future developments could include an assessment of the results by means of Monte Carlo simulations and feasibility analyses of aerocapture for different mission scenarios and on other planets, such as Venus.

## Appendix

The Mars Environment Multi-Model (MEMM) is a scientific tool that merges several existing models of the Martian environment:

- Mars-GRAM [24], an engineering-oriented model which provides a detailed description of the atmosphere of Mars, including temperatures, densities, pressures, wind speeds and chemical composition.
- MarsRAD [24], a companion tool of Mars-GRAM that evaluates the solar (short-wave) and thermal (long-wave) radiation fluxes at the surface and the top of the atmosphere.
- The model developed by Pollack [37] which provides the solar radiation fluxes at Mars and expands the results of MarsRAD by providing the direct and diffuse components of the short-wave radiation flux.
- The model due to Dycus [38] of the meteoroid flux at the top of the atmosphere of Mars based on experimental data.
- An improved version of the model by Dycus due to Divine [39] based on a more comprehensive set of experimental data.

MEMM integrates the above models into one tool. It computes and outputs the following physical parameters:

- atmosphere: average values and standard deviations of temperature, pressure, density, wind speed;
- properties of atmospheric constituents: concentration, specific heat, dynamic viscosity, heat convection coefficient;

- surface: ground (soil) temperature, local surface albedo, presence of polar ice, dust optical depth at ground, altitude of local surface with respect to the datum (defined as the elevation at which the atmosphere pressure is 610 Pa);
- positioning: Mars orbital radius, areocentric longitude of the Sun from Mars, local true solar time, solar zenith angle, azimuth and elevation of the Sun from a user-defined geographical location;
- radiation: long-wave flux and short-wave flux at the surface and the top of atmosphere, equivalent sky temperature, planetary albedo, atmosphere heating rates;
- meteoroids: flux at the top of the atmosphere.

MEMM has two main functionalities:

- it coordinates the execution of all the underlying models, which are hosted as independent modules but are executed as a single cooperative unit.
- it provides a common and consistent framework to automatically link the constituent models (i.e., the outputs of one model become the inputs of another). This is achieved by an extensive work of conditioning and organization of the inputs and outputs in a global pool of universally-defined physical quantities.

The models are coded in FORTRAN, whereas MEMM is a MATLAB script. The compiled binaries of the models are bridged into MATLAB via MEX modules. This choice provides a good computational performance while retaining the flexibility of MATLAB in interfacing, conditioning, pre- and post-processing.

In the context of the present contribution, MEMM significantly simplifies the operation, configuration and interfacing with Mars-GRAM.

**Funding Sources**

This work has been supported by Khalifa University of Science and Technology's internal grants FSU-2018-07 and CIRA-2018-85. J. Peláez and E. Fantino acknowledge also the support provided by the project entitled "Dynamical Analysis of Complex Interplanetary Missions," with reference ESP2017-87271-P sponsored by Spanish Agencia Estatal de Investigación (AEI) of Ministerio de Economía, Industria y Competitividad (MINECO) and by European Fund of Regional Development (FEDER). The work was also supported by the internal grants of the Department of Engineering of the University of Naples.